\documentclass[
preprint,
pra,
aps,
letterpaper,
groupedaddress
]{revtex4-2}

\usepackage{graphicx}   
\usepackage{longtable}
\usepackage{lineno,hyperref}
\usepackage{amsmath, amssymb}
\usepackage{array,multirow,bigdelim}
\begin{document}

	\title{Exactness of Semiclassical Quantization Rule for Broken Supersymmetry}
	
	\author{Asim Gangopadhyaya}
	\email{agangop@luc.edu}
	\author{Jonathan Bougie}
	\email{jbougie@luc.edu}
	\author{Constantin Rasinariu}
	\email{crasinariu@luc.edu}

	\affiliation {Department of Physics, Loyola University Chicago, Chicago, IL 60660, U.S.A}	
	\date{\today}
		
\begin{abstract}
	Semiclassical methods provide important tools for approximating solutions in quantum mechanics. In several cases these methods are intriguingly exact rather than approximate, as has been shown by direct calculations on particular systems. 
	In this paper we prove that the long-conjectured exactness of the supersymmetry-based
	semiclassical quantization condition for broken supersymmetry is a consequence of
	the additive shape invariance for the corresponding potentials.
\end{abstract}

\maketitle

\section{Introduction}

Semiclassical methods such as WKB provide an important tool to approximate  eigenvalues  of one-dimensional quantum mechanical problems, or the radial equation for systems with spherical symmetry \cite{Krieger67}. While these approximate values asymptotically converge to the exact values for large quantum numbers \cite{Dunham}, in particular cases these methods result in exact, rather than approximate spectra \cite{Bailey64,Krieger68,Krieger69,Bender, Sergeenko96}. Exact solutions are rare in quantum mechanics and hence of great interest when possible.  The conditions under which this occurs, if understood, could provide important insights into semiclassical methods.
	
In the context of Supersymmetric Quantum Mechanics (SUSYQM), the supersymmetric WKB approximation (SWKB) is exact for a class of potentials known as conventional potentials, for cases of unbroken supersymmetry \cite{Comtet}. This was recently shown to follow from the shape invariance of these potentials \cite{Gangopadhyaya2020}.
In this manuscript, we prove that the exactness of the modified broken supersymmetric WKB condition (BSWKB) conjectured almost three decades ago by Inomata et al. \cite{Inomata1,Inomata1a,Inomata2}, and Eckhardt \cite{Eckhardt},  follows directly from shape invariance of the corresponding conventional potential.
	
This paper is organized as follows. In subsection \ref{sec:SUSYQM}, we briefly review the basic tenets of SUSYQM and distinguish between broken and unbroken supersymmetry. We then present shape invariance in subsection \ref{sec:shapeinvariance} and conclude the section with an introduction to the BSWKB condition in \ref{sec:BSWKB}. In section \ref{sec:Methods}, we summarize previously derived results  and elaborate on the theoretical methods we employ to prove the exactness of BSWKB. Finally, we prove in section \ref{sec:Exact} that BSWKB exactness follows from shape invariance. 
	
\subsection{Supersymmetric Quantum Mechanics and Broken Supersymmetry}
\label{sec:SUSYQM}
A potential $V_-(x)$ in SUSYQM \cite{Witten,Solomonson,CooperFreedman,Bagchi,Cooper-Khare-Sukhatme,Gangopadhyaya-Mallow-Rasinariu,JunkerBook}
is related to a function $W(x)$, known as the superpotential, such that
 \begin{equation}
 V_-(x) = W^2(x)  -  \, \frac{\hbar}{\sqrt{2m}}\frac{dW(x)}{dx} ~.
 \end{equation}
We define two differential ladder operators
${\mathcal A}^\pm = \mp \, \frac\hbar{\sqrt{2m}}\frac{d}{dx}+ W(x)$ that are hermitian conjugates of each other.  Their product yields the hamiltonian  

\begin{equation}
H_-={\mathcal A}^+ {\mathcal A}^-  = \left(  -\hbar \frac{d}{dx}+ W(x) \right)  
\,\left( \hbar \frac{d}{dx}+ W(x)\right)
=  -\hbar^2 \frac{d^2}{dx^2}+ V_-(x) ~,
\label{A+A-}
\end{equation}
where we have set $2m=1$.
The product ${\mathcal A}^- {\mathcal A}^+$ yields a ``partner'' hamiltonian $H_+= -\hbar^2 \frac{d^2}{dx^2}+V_+(x)$, with potential $V_+(x) =
W^2(x) + \hbar \frac{d\, W}{dx}$. The two hamiltonians are intertwined by ${\mathcal A}^+ H_+ = H_- \, {\mathcal A}^+$ and ${\mathcal A}^- H_-
= H_+\,{\mathcal A}^-$. 

The eigenvalues $E^{\pm}_{n}$ cannot be negative since the hamiltonians $H_\pm$ are semi-positive definite. If either $E^{-}_{0}= 0$ or $E^{+}_{0}=0$, the system is said to have unbroken supersymmetry\footnote{Without loss of generality we can choose $E^{-}_{0}= 0$. If $E^{+}_{0}$ were equal to zero  instead, we could change $W\rightarrow -W$ and thus ensure that $E^{-}_{0} = 0$. Both $E^{-}_{0}$ and $E^{+}_{0}$ cannot be simultaneously zero.}. The intertwining of hamiltonians then yields the following relationships between their eigenvalues and eigenfunctions: 
\begin{eqnarray}
E^{-}_{n+1} =E^{+}_{n}, \quad \mbox{where}~ n=0,1,2,\cdots~, \label{eq:eigenvalueUB}
\end{eqnarray}
\begin{eqnarray}
\frac{~~~{\mathcal A}^- }{\sqrt{E^{+}_{n} }} ~\psi^{(-)}_{n+1} 
= ~\psi^{(+)}_{n} ~~, ~~\textrm{and} ~~
\frac{~~~{\mathcal A}^+}{\sqrt{E^{+}_{n} }}~\psi^{(+)}_{n} 
= ~  \psi^{(-)}_{n+1}~. \label{eq:isospectralityUB}
\end{eqnarray}
Additionally, $E^{-}_{0}=0$ leads to ${\mathcal A}^- \psi^{(-)}_{0}=0 $, and hence $\psi^{(-)}_{0}(x) \propto \exp\left[ -\frac{1}{\hbar}\int^x W(y) \, dy \right]$. Thus, the normalizability of the zero-energy groundstate requires that $\psi^{(-)}_{0}(\pm \infty) =0$; i.e.,  $\int^{\pm \infty} W(y) \, dy =\infty$, and hence the superpotential $ W(x)$ must have opposite signs at the left and right boundaries of the domain.

\begin{figure}[tbh]
	\centering
	\includegraphics[width=0.4\linewidth]{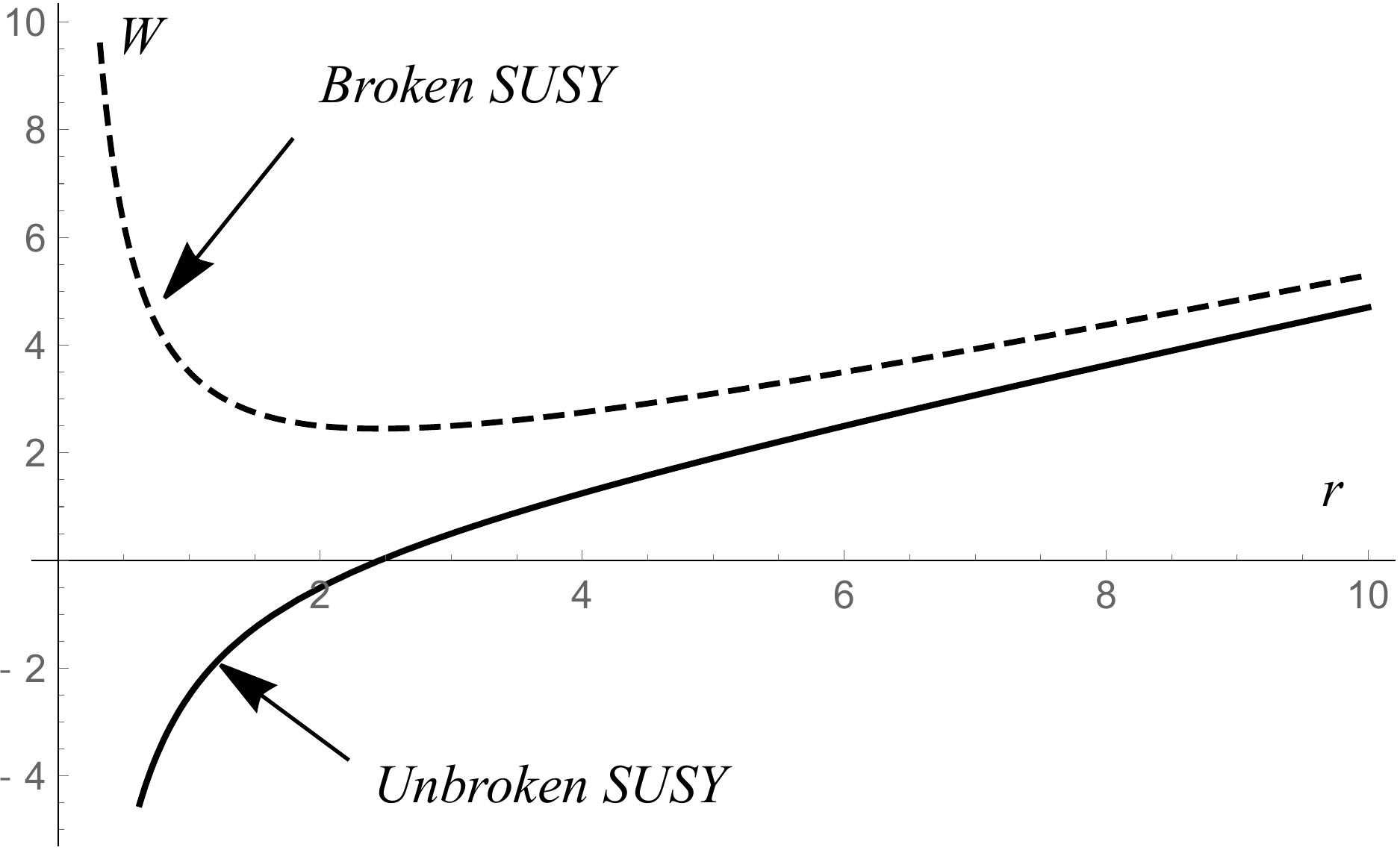}
	\caption{Broken ($\ell = -3$) vs. unbroken  ($\ell = 3$) phases of the 3-D oscillator superpotential. We have chosen units such that $\omega = 1$.}
	\label{fig:w-3d-o}
\end{figure}

If neither groundstate energy is zero, supersymmetry is broken. For broken SUSY, the superpotential $ W(x)$ must have the same sign at the left and right boundaries of the domain.
For example, the 3-D harmonic oscillator superpotential is given by 
\begin{equation}
	\label{eq:W3DO}
	W(r,\omega,\ell) = \frac{\omega r}{2} -\frac{\ell}r~~;\qquad  0 < r < \infty~.
\end{equation}
The supersymmetry is unbroken for $\ell > 0$ and broken for $\ell < 0$. 
In Fig. \ref{fig:w-3d-o} we illustrate the two phases. 
Hence the zero-energy groundstate for the broken phase is not normalizable, and $E^{-}_{0} = E^{+}_{0}>0$. In this case, the intertwining relations give 
\begin{eqnarray}
E_n^B\equiv E^{-}_{n} =E^{+}_{n}, \quad \mbox{where}~ n=0,1,2,\cdots~, \label{eq:eigenvalueB}
\end{eqnarray}
\begin{eqnarray}
\frac{~~~{\mathcal A}^- }{\sqrt{E^{+}_{n} }} ~\psi^{(-)}_{n} 
= ~\psi^{(+)}_{n} ~~, ~~\textrm{and} ~~
\frac{~~~{\mathcal A}^+}{\sqrt{E^{+}_{n} }}~\psi^{(+)}_{n} 
= ~  \psi^{(-)}_{n}~. \label{eq:isospectralityB}
\end{eqnarray}

\begin{figure}[tbh]
	\centering
	\includegraphics[width=\linewidth]{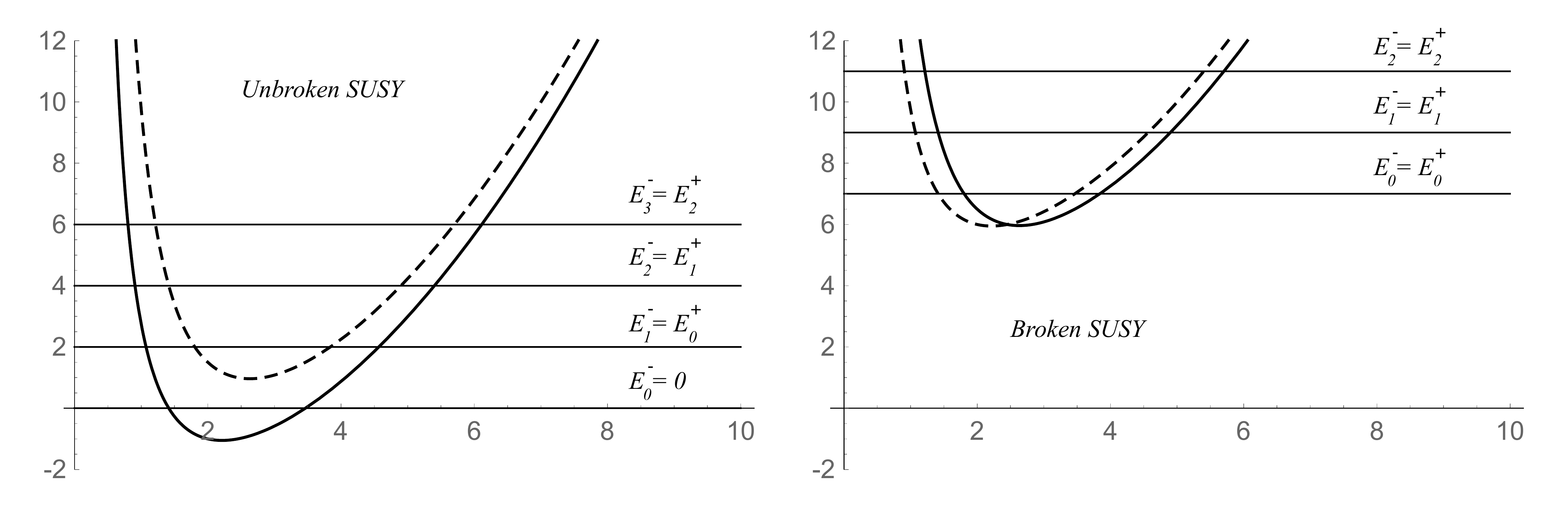}
	\caption{The potentials $V_+$ (dashed curve) and $V_-$ (solid curve) along with their eigenenergies for the 3-D oscillator for the unbroken ($\ell = 3$) and broken ($\ell = -3$) phases. We have chosen units such that $\hbar = \omega = 1$. }
	\label{fig:vpmb-ub}
\end{figure}

For example, the eigenvalues for the 3-D oscillator in its unbroken phase are the familiar $E_n^{-} = 2n\hbar\omega$, but in the broken phase they are given by \cite{Gangopadhyaya2001}  $E_n^B= (2n+1)\hbar\omega - 2\ell\omega$. In Fig.  \ref{fig:vpmb-ub}, we illustrate the 3-D oscillator potentials $V_{\pm}$ and their eigenvalues.

If we know the eigenvalues and eigenfunctions of one hamiltonian, then we can find the same for the partner hamiltonian using Eqs. (\ref{eq:eigenvalueUB} - \ref{eq:isospectralityUB}) in the unbroken phase, or  Eqs. (\ref{eq:eigenvalueB} - \ref{eq:isospectralityB}) in the broken phase. 
For superpotentials that have a property known as ``shape invariance,'' we can use the ladder operators to generate the entire spectra for a pair of partner potentials  in the unbroken supersymmetric phase, beginning from the  zero-energy groundstate of $V_-(x)$, as discussed in Sec. \ref{sec:shapeinvariance}.
	
\subsection{Shape Invariance}\label{sec:shapeinvariance}
A superpotential $W(x,a_i)$ is shape invariant
\cite{Infeld,Miller,gendenshtein1,gendenshtein2} if
\begin{equation}
W^2(x,a_i)  +  \hbar \frac{d\, W(x,a_i)}{dx}+g(a_i)=
W^2(x,a_{i+1})  -  \hbar \frac{d\, W(x,a_{i+1})}{dx}+g(a_{i+1})~ ,
\label{SIC1}
\end{equation}
for a set of parameters $a_i$, where the relationship between parameters is given by a function $f(a)$ such that $a_{i+1} = f({a_i})$ and the function $g(a)$ gives the difference between energy levels. We then have
\begin{eqnarray}
E_n^{(+)}(a_0)-E_n^-(a_1)&=&g(a_1)-g(a_0),\label{eq:eigenvalues} \\
\psi^{(+)}_{n}(x,a_0)&=&
\psi^{(-)}_n(x,a_1) \label{eq:eigenfunctions}
~.
\end{eqnarray}

For shape invariant superpotentials with unbroken supersymmetry, the existence of a groundstate with $E^{(-)}_{0}= 0$ allows us to use Eqs.(\ref{eq:eigenvalues} , \ref{eq:eigenfunctions}) to construct the entire spectra of $H_-$ and $H_+$ from this ground state. For unbroken SUSY, therefore,
\begin{eqnarray}
	E_n^{(-)}(a_0)&=&g(a_n)-g(a_0), \label{eq:En} \\
	\psi^{(-)}_{n}(x,a_0)&=&
	\frac{{\mathcal A}^+{(a_0)} 
		~ {\mathcal A}^+{(a_1)}  \cdots  {\mathcal A}^+{(a_{n-1})}}
	{\sqrt{E_{n}^{(-)}(a_0)\,E_{n-1}^{(-)}(a_1)\cdots E_{1}^{(-)}(a_{n-1})}}~\psi^{(-)}_0(x,a_n)
	~.
\end{eqnarray}
Therefore shape invariance leads directly to the exact solvability of quantum mechanical systems. This result reflects the fact that shape invariance is a symmetry condition intrinsically connected to an underlying potential algebra \cite{Fukui1993, asim1, balantekin1, asim2, asim3,  balantekin2}.

In this paper we consider two types of shape invariance: an additive shape invariance described by $a_{i+1}=a_{i}+\hbar$, and a discrete shape invariance, to be described later, that leads to a change of phase between broken and unbroken supersymmetry. As shown in Ref. \cite{Gangopadhyaya2001}, the combination of such shape invariances allows us to determine the spectrum of additive shape invariant systems in the broken SUSY phase.

Additive shape invariant superpotentials $W(x,a_i)$ that do not depend on $\hbar$ are called \cite{Bougie2010, symmetry} ``conventional.''\footnote{Several authors \cite{Ramos00,Ramos99} have called these potentials ``Classical''.}
They satisfy the following set of partial differential equations: 
\begin{eqnarray}
	W \, \frac{\partial W}{\partial a} - \frac{\partial W}{\partial x} + \frac12 \, \frac{d g(a)}{d a}= 0~, \label{PDE1} \\
	\frac{\partial^{3}}{\partial a^{2}\partial x} ~W(x,a)= 0~.\label{PDE2}
\end{eqnarray}
This idea was first noted in Ref. \cite{Gangopadhyaya2008} and fully developed in Refs. \cite{Bougie2010,symmetry}, where it was demonstrated that Ref. \cite{Dutt_SUSY} provided the complete list of conventional superpotentials.
Additional ``extended'' shape invariant superpotentials have been found \cite{Quesne1,Quesne2,Quesne2012a,Quesne2012b,Odake1,Odake2,Tanaka,Odake3,Odake4,Ranjani1}, that each consist of a conventional superpotential along with an additive extension that depends explicitly on $\hbar$ \cite{Bougie2010,symmetry}.

\subsection{The BSWKB Condition}
\label{sec:BSWKB}

The following supersymmetric WKB (SWKB) condition was proposed in 1985 by Comtet et al. \cite{Comtet} 
\begin{equation}
\int_{x_1}^{x_2} \sqrt{E_n-W^2(x,a)}\,\, \textrm{d}x =  n \pi\hbar~, \quad \mbox{where}~ n=0,1,2,\cdots \label{eq:swkb}~,
\end{equation}
where $E_n$ is the energy of the system and the limits ${x_1}$ and ${x_2}$ are  solutions of $W(x,a)=\pm \sqrt{E_n}$. The exactness of SWKB for all conventional potentials was recently proven to follow from shape invariance for cases in which supersymmetry is unbroken \cite{Gangopadhyaya2020}.

In the 1990s, Inomata and Junker \cite{Inomata1, Inomata1a, Inomata2,JunkerBook}, and Eckhardt \cite{Eckhardt} proposed a modified SWKB condition for broken supersymmetry (BSWKB):
\begin{eqnarray}
\int_{x_1}^{x_2} \sqrt{E_n^{B}-W^2(x,a)}\,\, \textrm {d}x =  \left( n+\frac12\right)  \pi\hbar~, \quad \mbox{where}~ n=0,1,2,\cdots \label{eq:bswkb}~.
\end{eqnarray}
In this paper we prove that Eq.(\ref{eq:bswkb}) follows from additive shape invariance of the unbroken phase for all conventional superpotentials that have two solutions of  $W(x,a)=\pm \sqrt{E_n^{B}}$ in their broken phase so that the limits of the integral are properly defined. 

\section{Methods}
\label{sec:Methods}
In this section, we elaborate on the theoretical methods we will use to prove the exactness of BSWKB, Eq. (\ref{eq:bswkb}).
We begin by defining a function $I(a,n,\hbar)$ by
\begin{eqnarray}
I(a,n,\hbar) \equiv
\int_{x_1}^{x_2} \sqrt{E_n^B-W^2(x,a)}\,\, \textrm{d}x  = \frac12 \oint \sqrt{E_n^B-W^2(x,a)}\,\, \textrm{d}x
\label{eq:swkb1}~,
\end{eqnarray}
where $E_n^B$ represents the energy in the broken phase, and the last integral is carried out on the complex $x$-plane enclosing the cut from ${x_1}$ to ${x_2}$. 

Unlike the unbroken case, for broken SUSYQM, the groundstate energy $E_0^B$ is unknown. Therefore, additive shape invariance combined with isospectrality of the partner hamiltonians given by Eqs. (\ref{eq:eigenvalueUB}) and (\ref{eq:isospectralityUB}) is not sufficient to determine $E_n^B$.

\begin{figure}[h!]
	\centering
	\includegraphics[width=0.5\linewidth]{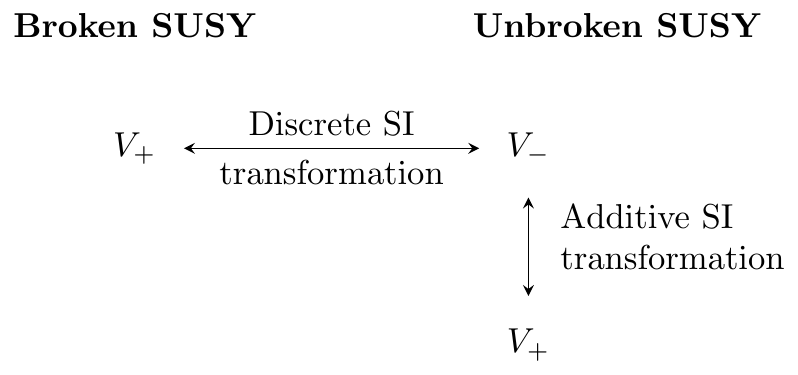}
	\caption{Schematic illustrating the transformations used to determine the spectra for the systems in their broken phases.}
	\label{fig:transformations}
\end{figure}

Hence, following Ref. \cite{Gangopadhyaya2001}, we first carry out a change of parameter that takes the system from the broken to an associated unbroken phase, and then use additive shape invariance to determine its spectrum. We then use the isospectral relations of Eqs. (\ref{eq:eigenvalueUB}) and (\ref{eq:isospectralityUB}) to determine the eigenvalues and eigenfunctions for the broken phase, as illustrated by Fig. \ref{fig:transformations}.

\medskip
To show the exactness of BSWKB for conventional superpotentials, we now divide the conventional potentials into three classes and use properties of each class that were identified in \cite{Gangopadhyaya2020}. We then examine the integral $I$ from Eq.~\ref{eq:swkb1} for each class. Since these classes are exhaustive for conventional superpotentials, we thus prove that the BSWKB condition is exact for all cases in which the integral is properly defined. 

The general solution \cite{Bougie2010,symmetry} to Eq. (\ref{PDE2}) is
\begin{equation}
W(x,a)= a\,f_1(x) +f_2(x)+u(a) \label{eq:superpotential} ~.
\end{equation}
This form was suggested by Infeld et al. \cite{Infeld},  further studied in \cite{Ramos99,Ramos00}, and proven as the general solution in \cite{Bougie2010, symmetry, Gangopadhyaya2020}.

Since both $f_1(x)$ and $f_2(x)$ cannot be simultaneously constants (otherwise $W(x,a)$ would be a constant), the following three classes encompass all possible conventional superpotentials. Class I: $f_1=\alpha$, a constant; Class II: $f_2=\mu$, a constant;  Class III: $f_1$ and $f_2$ are both functions of  $x$. In Table \ref{table:classes}, we list key properties for each of these classes that were shown in \cite{Gangopadhyaya2020} to follow from their additive shape invariance.
\vspace*{0.2in}

\begin{table*} [h!]
	\begin{center}
		\begin{ruledtabular}
		\begin{tabular}{lllrll}
	
			Class&Form of $W$ & Constraints from && \hspace{-0.2cm}Subclasses&$E_n^{-}$ for \\ 
			&&shape invariance &&&unbroken SUSY\\ \hline
			\multirow{2}{*}{Class I}& \multirow{2}{*}{$f_2(x)+\alpha\, a$} & \multirow{2}{*}{$\alpha f_2-f_2'=\varepsilon \equiv -\omega/2$ }& \multirow{2}{*}{$\Biggl\lbrace$}&\hspace{-0.15cm}IA: $\alpha=0$&$n\omega\hbar$\\
			&&&&\hspace{-0.15cm}IB: $\alpha\neq0\hspace{1cm}$  &$\alpha^2\,a^2 -  \alpha^2 (a+n\hbar)^2$\vspace{0.5cm}\\
			\multirow{2}{*}{Class II}& \multirow{2}{*}{$a f_1(x)+B/a$} & \multirow{2}{*}{$f_1^2-f_1'=\lambda$}&
			\multirow{2}{*}{$\Biggl\lbrace$}&\hspace{-0.15cm}IIA: $\lambda=0$& $\frac{B^2}{a^2} - \frac{B^2}{(a+n\hbar)^2} 
			$\\
			&&&&\hspace{-0.15cm}IIB: $\lambda\neq0$&$\frac{B^2}{a^2} - \frac{B^2}{(a+n\hbar)^2} +\lambda\left[\, a^2-(a+n\hbar)^2\right]
			$\vspace{0.5cm}\\
			\multirow{2}{*}{Class III}& \multirow{2}{*}{$a f_1(x)+f_2(x)$} &$f_1^2-f_1'=\lambda$,&
			\multirow{2}{*}{$\Biggl\lbrace$} &\hspace{-0.15cm}IIIA:$\lambda=0$ &$2n\omega \hbar$\\
			&&$f_1f_2-f_2'=\varepsilon\equiv-\omega$&&\hspace{-0.15cm}IIIB:$\lambda\neq0$&$\lambda\left[a^2-(a+n\hbar)^2\right]$\\
		\end{tabular}
	\end{ruledtabular}
		\caption{Three classes of conventional shape invariant superpotentials and their properties as shown in \cite{Gangopadhyaya2020}. The following are all constants: $\alpha$, $\varepsilon$, $\omega$, $\lambda$, and $B$.}
		\label{table:classes}
	\end{center}
\end{table*}

\section{The Exactness of BSWKB}
\label{sec:Exact}

At this point, we are positioned to prove the exactness of BSWKB. Since Classes I-III represent all possible conventional superpotentials, we will examine each class separately. 

\subsection{Class I}
For this class, $W(x,a)=f_2(x)+\alpha\,a$, where $f_2' = \alpha f_2 - \varepsilon$. For $\alpha = 0$, $f_2' = -\varepsilon$ and for $\alpha \ne 0$, $\varepsilon = 0$, so $f_2^\prime = \alpha f_2$.
\footnote{For $\alpha\ne 0$,  a particular solution for  $f_2$ is $\varepsilon/\alpha$. By redefining $a+\varepsilon/\alpha \rightarrow  a$, we can set  $\varepsilon=0$.}
In either case, $W'=f_2'$ cannot cross zero and therefore has a definite sign. 
Hence $W$ is a monotonic function.  Since for the broken SUSY phase the left and right limits must have the same signs, we see that $W(x)$ is never zero. 

This then implies that $(W^2)' = 2 W W'$ is monotonic and cannot vanish at any point, and hence $W^2(x)$ has no minimum within the domain.  Consequently, $E_n^B=W^2$ has only one intersection point, not two.

\subsection{Class II}

For this class, $W(x,a)=a f_1(x)+B/a$ and $f_1'=f_1^2-\lambda$. Thus $W^\prime(x,a)=a f_1'(x).$ For $\lambda=0$,  $f_1'= f_1^2>0$. For $\lambda<0$, $f_1^\prime=f_1^2+|\lambda|>0$. For $\lambda>0$, we must have either $(f_1)^2>\lambda$ or $(f_1)^2<\lambda$, as $f_1^2 \ne \lambda$ at any point in the domain, or it would yield a constant $W$ which is a trivial solution. Hence in all cases $W'=f_1'$ has a fixed sign and $W$ is a monotonic function.  Since in the broken phase $W$ must have same signs at both ends, it cannot have a zero within the domain of $x$.  Consequently, $E_n^B=W^2$ has only one intersection point, not two.

\subsection{Class III}
For this class, $W(x,a)=a f_1(x) + f_2(x)$. Here $f_1^\prime=f_1^2 -\lambda$ and $f_2^\prime = f_1 f_2- \varepsilon$, where $\lambda$ and $\varepsilon$ are constants. 
In this case, since $W$ depends on both $f_1$ and $f_2$, it could have a minimum and therefore $W^2=E_n^B$ could indeed have two intersection points. This class further splits into two cases:  $\lambda= 0$ and $\lambda\neq 0$. 
Next we examine each of these two cases separately and prove that the BSWKB condition is exact. 
\subsubsection{Class IIIA: $\lambda =0$}

For this subclass $W=a f_1 + f_2$ and $\lambda = 0 $.  
Therefore $f_1'=f_1^2$, and $f_1$ must be nonzero everywhere.
The function $f_2$ satisfies 
\begin{equation}
	\label{eq:f2}
	f_2'=f_1\,f_2 -\varepsilon~.
\end{equation}
The homogeneous equation of (\ref{eq:f2}) is solved by $f_2 = \gamma f_1$, which can be absorbed into the first term of $W$ with a redefinition of $a$. A particular solution of (\ref{eq:f2}) is $f_2 = \frac12 \varepsilon/f_1$, so the superpotential reduces to $W=a f_1+ \frac12 \varepsilon/f_1$. Since the BSWKB integral (\ref{eq:swkb1}) is invariant under $W \to -W$, we can choose the sign of $\varepsilon$ without loss of generality.  Substituting  $W=a f_1+ \frac12 \varepsilon/f_1$ into Eq. (\ref{PDE1}), we get $dg/da=-2\varepsilon$ which should be positive to avoid level crossing. Therefore we choose $\varepsilon < 0$. 

If $a<0$, $W$ will never cross zero, so we must have broken SUSY. However, if $a>0$, $W$ will have opposite signs as $f_1\to 0$ and $|f_1|\to\infty$, so this corresponds to unbroken SUSY.  
Therefore, the potential in its broken phase corresponds to its counterpart in the unbroken phase through the discrete shape invariance transformation $a$ to $-a$, which we can use to find the energy spectrum in the broken phase. 

The partner potentials are given by
	\begin{eqnarray}
		V_\pm (x,a) &=& a^2 f_1^2  \, + \, \frac14 \varepsilon^2/f_1^2  \, + \,  a\varepsilon ~\pm \,
		\hbar\,  \left( af_1^2- \frac12 \varepsilon\right) 
		\nonumber\\
&=& a(a\pm \hbar) f_1^2 \, + \,\frac14 \varepsilon^2/f_1^2  \, + \, a\varepsilon \, \mp \,
		\frac12 \hbar \varepsilon~.
	\end{eqnarray}
They are shape invariant under two distinct parameter changes
\begin{itemize}
	\item Additive Shape Invariance:  $V_+(x,a) - V_-(x,a+\hbar) = -2 \hbar \varepsilon $ ,
	\item Discrete Shape Invariance: $V_+(x,a) - V_-(x,-a) = (2 a - \hbar) \varepsilon $ .
\end{itemize}
Since the discrete shape invariance transforms the broken into the unbroken phase and the energy for the unbroken phase is $g(a+n\hbar) - g(a)= -2 \hbar n \varepsilon$,  the energy in the broken phase is given by
	$E_n^B= (2 a - \hbar) \varepsilon -2 \hbar n \varepsilon = \left( 2 a - \hbar (1 + 2 n)\right)   \varepsilon$.
	
\begin{figure}[h!]
	\centering
	\includegraphics[width=0.45\linewidth]{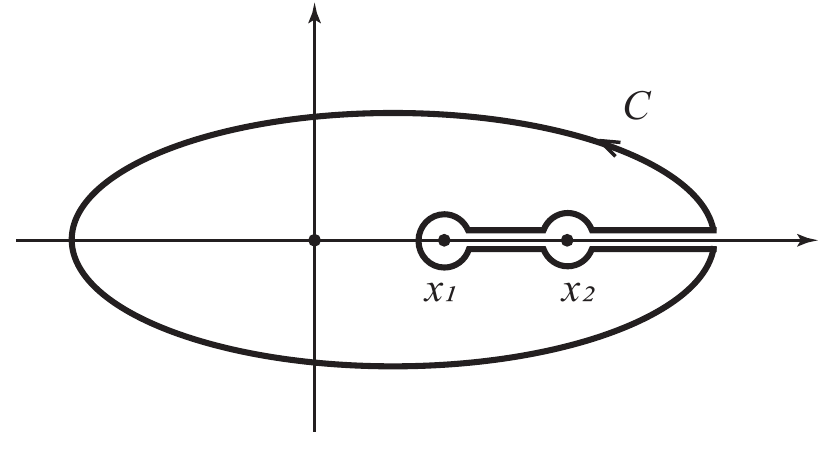}
	\caption{Complex integration. Since the integration contour $C$ travels across $x_1$ and $x_2$ twice,
	$I(a,n,\hbar) \equiv \int_{x_1}^{x_2}
	\sqrt{E_n^B-W^2} {~dx} = \frac 12\oint \sqrt{E_n^B-W^2} {~dx} $. 
 }
	\label{fig:z0}
\end{figure}

	We will now compute 
	\begin{eqnarray}I(a,n,\hbar) \equiv \int_{x_1}^{x_2}
		\sqrt{
			E_n^B-W^2} {~dx} = \int_{x_1}^{x_2}
		\sqrt{
			E_n^B-\left(af_1+ \frac12 \varepsilon/f_1
			\right)^2} {~dx}.
	\end{eqnarray}	
To calculate the integral, we move to the complex $x$-plane, as illustrated in Fig. \ref{fig:z0}. 

\noindent
We obtain  
\begin{eqnarray}
	I(a,n,\hbar) 
	&=&
	\frac12 \oint \frac{\sqrt{
			4E_n^B\,f_1^2-\left(2af_1^2+  \varepsilon
			\right)^2} {~dx}}{2f_1}
	= 
	\frac14 \oint \frac{\sqrt{
			4E_n^B\,f_1^2-\left(2af_1^2+  \varepsilon
			\right)^2} {~df_1}}{f_1 f_1^\prime} 
	\nonumber\\
	&=& \frac18 \oint \frac{\sqrt{
			4E_n^B\,f_1^2-\left(2af_1^2+  \varepsilon
			\right)^2} {~\left( 2f_1 df_1\right) }}{f_1^4} ~.  
\end{eqnarray}
Setting ${f_1^2=z}$, we get
\begin{eqnarray}
I(a,n,\hbar) 	&=& \frac18 \oint \frac{\sqrt{
			4E_n^B\,z-\left(2az+  \varepsilon
			\right)^2} {~dz}}{z^2}~.
		\label{eq:class3a-divergence}
\end{eqnarray}
In addition to having a second order pole at $z=0$, 
the expression inside the integral sign approaches  $\left| z\,dz/z^2\right| = d\theta$ for large values of $|z|$. Hence, the contribution from the outer circle $I_\infty$, generated by a singularity at infinity, is nonzero. The integral gives
\begin{eqnarray}
I(a,n,\hbar) 	&=& \frac{2\pi i}8  \left. \left( \frac{\partial}{\partial z} \, \sqrt{
		4E_n^B\,z-\left(2az+  \varepsilon
		\right)^2}\right) \right|_{z \rightarrow 0} +  I_\infty
	\nonumber\\
	&=& -\frac{1}{2} \pi  (a-h (2 n+1))
 + I_\infty	 
 \nonumber\\
 &=& \left( n+\frac12\right) \pi \hbar -\frac{1}{2} \pi a + I_\infty. \label{eq:class3a-divergence2}
\end{eqnarray}
To determine $I_\infty$, we set $t=1/z$ in 	Eq.~\ref{eq:class3a-divergence}.  This gives
\begin{eqnarray}
I_\infty
	&=&\frac18 \oint\frac{\sqrt{-4 a^2+4 a\, t\, \varepsilon -4 E_n^B t+t^2 \varepsilon ^2} }{t}\quad dt
\nonumber\\
	&=&\frac{2\pi i}8 \left. 
	\sqrt{-4 a^2+4 a\, t\, \varepsilon -4 E_n^B t+t^2 \varepsilon ^2} \right|_{t\rightarrow 0}
\nonumber\\
	&=& \frac{1}{2} \pi a ,
	\label{eq:class3a-divergence3}
\end{eqnarray}
where we have used $a<0$ for the broken phase.  Combining results of Eqs. (\ref{eq:class3a-divergence2}) and (\ref{eq:class3a-divergence3}), we get
$$ 	I(a,n,\hbar) = \left( n+\frac12\right) \pi \hbar -\frac{1}{2} \pi a + \frac{1}{2} \pi a = \left( n+\frac12\right) \pi \hbar~.$$

\subsubsection{Class IIIB: $\lambda \neq 0$}
In this case  we have $W(x,a)=a f_1(x) + f_2(x)$, where $f_1^\prime=f_1^2 -\lambda$ and $f_2^\prime = f_1 f_2- \varepsilon$, where $\lambda$ and $\varepsilon$ are constants. Without loss of generality (WLOG), we introduce a parameter $B$ which is independent of $a$, so that  $f_2 \to B f_2$.  Thus, the partner potentials are given by
\begin{widetext}
\begin{eqnarray}V_\pm (a,B,r) &=&  \left(a^2 \pm a \hbar +B^2\right){f_1}^2+B  (2 a \pm \hbar) {f_1}(x) {f_2}(x)
	+\lambda  \left(\mp\, a \hbar -B ^2\right) \nonumber\\
	&=&\left[ a(a\pm \hbar) + B^2\right] \, {f_1}^2
	+2B  \left( a \pm \hbar/2 \right)  {f_1}(x) {f_2}(x)
	+\lambda  \left(\mp\, a \hbar -B ^2\right)~.
\end{eqnarray}
\end{widetext}
In addition to additive shape invariance generated by $a\rightarrow a+\hbar$, these two potentials also satisfy a phase changing discrete shape invariance via $a \rightarrow B+\hbar/2, B\rightarrow a+\hbar/2$ as shown in the Appendix.  \footnote{Previously, alternate parametric transformations have been used in other contexts  \cite{Ghosh98,Ramos2000b, Yadav2016}.} That is,
\begin{itemize}
	\item Additive Shape Invariance:  $V_+(a,B,r) - V_-(a+\hbar,B,r)  =  \lambda\left[a^2-(a+\hbar)^2\right]$ ,
	\item Discrete Shape Invariance: $V_+(a,B,r) - V_-(B+\frac{\hbar}{2}, a+\frac{\hbar}{2},r) =   {\lambda  \left(a^2-(B +\frac{\hbar}{2})^2\right)}$ .
\end{itemize}

We now use the corresponding unbroken phase to find the spectrum in the broken phase. From additive shape invariance for the potential $V_-(a,B,r)$ with unbroken SUSY,  the energy eigenvalues are given by $E_n=\lambda\left[a^2-(a+n\hbar)^2\right]$. The proper hierarchy of eigenvalues: $E_{n+1}>E_n>E_{n-1}$ requires
\begin{equation}
	\lambda(a+n\hbar)<0 ~ \label{eq:Condition4a}~.
\end{equation}
Let $E_n^B$ be the energy for the system with potential $V_+ (a,B,r)$ with broken SUSY, and  $E_n$ be the energy for $V_-(B+\hbar/2, a+\hbar/2,r)$ with unbroken SUSY. Then from the discrete shape invariance condition we have 
\begin{eqnarray}E_n^B&=&E_n+ {\lambda  \left[a^2-\left(B +\frac{\hbar}{2}\right)^2\right]}
	\nonumber\\
	&=& \lambda\left[(B+\frac{\hbar}{2})^2-(B+\frac{\hbar}{2}+n\hbar)^2\right]+ {\lambda  \left[a^2-(B +\frac{\hbar}{2})^2\right]}
	\nonumber\\
	&=& \lambda  \left[a^2-{(B +n\,\hbar+\frac{\hbar}{2})^2}\right].
\end{eqnarray}

The next step in computing the BSWKB integral is to note that the
homogeneous and particular solutions of $f_2^\prime = f_1 f_2- \varepsilon$ are  $\sqrt{\left| f_1^2-\lambda\right| }$  and $\frac{\varepsilon}{\lambda}\,f_{1}$, respectively \footnote{The homogeneous solution is
	$
	f_2 	\propto \exp{\left[ \int f_1 \, dx\right]}  = \exp{\left[ \int f_1 \, \frac{df_1}{f_1^2-\lambda}\right]}  = \exp{\left[ \frac12  \int \frac{df_1^2}{f_1^2-\lambda}\right]}  = \sqrt{\left| f_1^2-\lambda\right| }
	$
}. 
Thus, a redefinition of the parameter $a$ yields $W= af_1+B \sqrt{\left| f_1^2-\lambda\right| }$.

Since\footnote{Since $f_1^\prime=f_1^2 -\lambda$, if $f_1^2$ were to equal $\lambda$ at any point, then all derivatives of $f_1$ would vanish. In this case, we would get $f_1^2 =\lambda$ for all values of $x$, which would result in a trivial $W$.}  $f_1^2 \neq \lambda$, we follow a similar procedure as \cite{Gangopadhyaya2020} and introduce a function
	$y (x)\equiv \frac{\sqrt{\lambda} -f_1}{\sqrt{\lambda}+f_1}$. Thus, $y'=2y\,\sqrt{\lambda}$ and $f_1 = \sqrt{\lambda}\left(\frac{y-1}{y+1} \right) $. Then the functions ${\mathcal{S}}(x)\equiv\frac{y^{1/2}-y^{-1/2}}{2\sqrt{\lambda}}$, and ${\mathcal{C}}(x)\equiv\frac{y^{1/2}+y^{-1/2}}{2}$ satisfy the identities: 
	$$
	\begin{array}{lll}
		{d{\mathcal{C}}}/{dx}=\lambda {\mathcal{S}}~, & {d{\mathcal{S}}}/{dx}={\mathcal{C}}~, &
		{\mathcal{C}}^2(y)-\lambda \,{\mathcal{S}}^2(y) = 1~, \\
		2\,{\mathcal{C}}(y)\,{\mathcal{S}}(y) = {\mathcal{S}}(y^2)~, \quad \quad &  {\mathcal{C}}^2(y)+\lambda \,{\mathcal{S}}^2(y) = {\mathcal{C}}(y^2)~.
	\end{array}
	$$
	Writing $f_1$ and $f_2$ in terms of $\mathcal{S}$ and $\mathcal{C}$ yields $f_1 = -\lambda \frac{{\mathcal{S}}}{{\mathcal{C}}}$ and  $f_2 = B \sqrt{f_1^2-\lambda} =  \frac{B}{{\mathcal{C}}}$. Thus, 
the superpotential becomes
$$W(x,a) = -a~\frac{\lambda {\mathcal{S}}}{{\mathcal{C}}}+\frac{B}{{\mathcal{C}}} ~.$$

\begin{figure}[h!]
	\centering
	\includegraphics[width=0.4\linewidth]{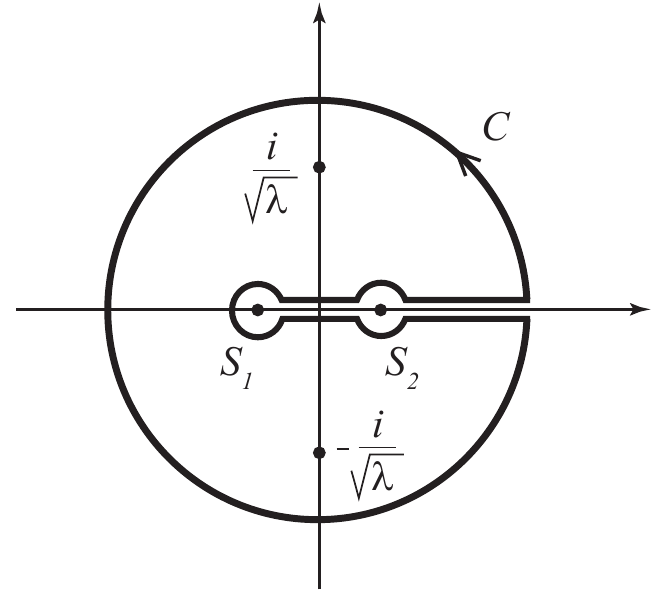}
	\caption{BSWKB complex integration. $S_1$ and $S_2$ are the zeroes of the numerator of Eq. (\ref{eq:ClassIIIbIntegral1}).}
	\label{fig:-1}
\end{figure}

\noindent
We now compute the BSWKB integral
\begin{eqnarray}
	I(a,n,\hbar)
	&=& \int_{x_1}^{x_2} {\sqrt{E_n^B-\left( -\frac{\lambda {\mathcal{S}} a}{{\mathcal{C}}}+\frac{B}{{\mathcal{C}}}\right) ^2}} ~ {dx}
=\frac12 \oint {\sqrt{E_n^B-\left( -\frac{\lambda {\mathcal{S}} a}{{\mathcal{C}}}+\frac{B}{{\mathcal{C}}}\right) ^2}} ~ {dx} ~.	\nonumber
\end{eqnarray}
After some algebra, we get
\begin{eqnarray}
	I(a,n,\hbar)	
	&=&\frac12 \oint \frac{\sqrt{E_n^B\, \left(1+ \lambda{\mathcal{S}}^2\right) -\lambda^2a^2{\mathcal{S}}^2+2\lambda {\mathcal{S}}B a-B^2}}{\left(1+ \lambda{\mathcal{S}}^2\right)}
	{d{\mathcal{S}}}~, \label{eq:ClassIIIbIntegral1}
\end{eqnarray}	
where we used $\frac{dx}{\mathcal{C}} = \frac{d{\mathcal{S}}}{{\mathcal{C}}^2} = \frac{d{\mathcal{S}}}{\left(1+ \lambda{\mathcal{S}}^2\right)}$. The integrand in Eq. (\ref{eq:ClassIIIbIntegral1}) 
has poles at ${\mathcal{S}} = \pm \frac{i} {\sqrt{\lambda}}$ and has a singularity at infinity as shown in Fig \ref{fig:-1}. Hence, the integral is given by

\begin{eqnarray}
	I(a,n,\hbar)
	&=&\frac{(2\pi i) }2 
	\left.  \frac{\sqrt{E_n^B\, \left(1+ \lambda{\mathcal{S}}^2\right) -\lambda^2a^2{\mathcal{S}}^2+2\lambda {\mathcal{S}}B a-B^2}}{\left(1+ i \sqrt{\lambda}{\mathcal{S}}\right)} \right|_{{\mathcal{S}} \rightarrow \frac{-i}{\sqrt{\lambda}}} 
	\nonumber\\
	&+&	\frac{(2\pi i) }2 \left.  \frac{\sqrt{E_n^B\, \left(1+ \lambda{\mathcal{S}}^2\right) -\lambda^2a^2{\mathcal{S}}^2+2\lambda {\mathcal{S}}B a-B^2}}{\left(1- i \sqrt{\lambda}{\mathcal{S}}\right)} \right|_{{\mathcal{S}} \rightarrow \frac{i}{\sqrt{\lambda}}} \quad 
	+ \quad I_\infty
	\nonumber\\
	&=&  -\frac{1}{2} \pi  \left(B +i a \sqrt{\lambda }\right)   
	-\frac{1}{2} \pi  \left(B -i a \sqrt{\lambda }\right) 
		+ I_\infty \label{eq:ClassIIIbIntegral2}
\end{eqnarray}
To determine $I_\infty$, we set $t= 1/{\mathcal{S}}$ and ${d\mathcal{S}}= -dt/t^2$  in Eq. (\ref{eq:ClassIIIbIntegral1}).
\begin{eqnarray}
I_\infty 	
	&=& -\, \frac12 \oint \frac{\sqrt{{E_n^B} \left(\lambda +t^2\right)-(B  t-a \lambda )^2}}{t\left( \lambda +t^2\right) } ~dt
	\nonumber\\
	&=&  - \frac{(-2\pi i)}2 \left. 
	\frac{\sqrt{{E_n^B} \left(\lambda +t^2\right)-(B  t-a \lambda )^2}}{\left( \lambda +t^2\right) }\,  \right|_{t\rightarrow 0}\nonumber\\
	& = &  \frac{\pi i}{\lambda} \left. 
	\sqrt{{E_n^B} \lambda -(a \lambda )^2}\,  \right|_{t\rightarrow 0}
	\nonumber\\ 
	&=&   \pi   \left( B +n \hbar +\hbar/2 \right) ~,\label{eq:ClassIIIbIntegral3}
\end{eqnarray}
where  \footnote{Here we replace $a$ by $B +\hbar/2 $ in Eq. (\ref{eq:Condition4a}).} we have used $\lambda \left( B + n \hbar +\hbar/2 \right)<0$. The negative sign of the second equality comes from the clockwise contour. Collecting contributions from Eqs. (\ref{eq:ClassIIIbIntegral2}) and (\ref{eq:ClassIIIbIntegral3}), we find 
$$	I(a,n,\hbar) =  -\frac{1}{2} \pi  \left(B +i a \sqrt{\lambda }\right)   
-\frac{1}{2} \pi  \left(B -i a \sqrt{\lambda }\right) +  \pi   \left( B +n \hbar +\hbar/2 \right)  = \left( n+\frac12 \right) \pi \hbar ~.
$$

\section{Conclusion}

In conclusion, we proved that the BSWKB semiclassical quantization condition is exact, and that this exactness follows from the underlying additive shape invariance of the systems. 

Our work differs from previous results because we did not tie our calculations to any particular case. Instead, we used the general form of the superpotential $W$  arising from the additive shape invariance condition for conventional potentials to arrive at our result. Fundamental to our approach was the connection between the unbroken and broken phases of SUSY which allowed us to find the energy spectrum and to compute the BSWKB integral for the broken phase.

In Ref. \cite{Gangopadhyaya2020}, the authors demonstrated that the exactness of SWKB results from the additive shape invariance condition for conventional potentials in the unbroken phase. In this paper we prove that BSWKB exactness is a consequence of the same condition for the broken phase. Together, these papers highlight the essential role played by the additive shape invariance and related algebraic symmetries in both phases. These symmetries are relevant not only to the exactness of these semiclassical methods, but also to the exact solvability of quantum mechanical systems \cite{Fukui1993, asim1, asim2, asim3, balantekin1, balantekin2}.

\appendix*
\section{Changes of parameters and corresponding phase transformations for Class IIIB ($\lambda \neq 0$)}

In this case, we have $W(x,a)=a f_1(x) + f_2(x)$, where $f_1^\prime=f_1^2 -\lambda$ and $f_2^\prime = f_1 f_2$ \footnote{Without loss of generality, we have set $\varepsilon=0$. See Ref. \cite{Gangopadhyaya2020} for details.} This yields $f_2=B\sqrt{f_1^2-\lambda}$ and $W=af_1+B\sqrt{f_1^2-\lambda}$. Note that $f_1^2$ cannot equal $\lambda$ at any point in the domain, or all derivatives of $f_1$ would be zero and $W$ would be a constant.
This case breaks up in three sub-cases:
\begin{enumerate} 
	\item \label{Enmurate:lamda<0} $\lambda<0$. 
	
	In this case $f_1^\prime = f_1^2+ \left| \lambda\right| \ge \left| \lambda\right| $. Therefore, $f_1$ is monotonic and unbounded.
	Let $x_0$ be the point where  $f_1=0$.  The derivative of  $f_1$ increases to the left and right of $x_0$. Furthermore, at $x_0$, all even derivatives are zero and all odd derivatives are positive. Thus $f_1$ is antisymmetric about $x_0$. Since $f_2  = B \sqrt{f_1^2+ \left| \lambda\right|} \rightarrow B \left| f_1(x)\right|$ as $x\rightarrow x_L \mbox{ or }x_R$, broken SUSY requires $a \le B$ so that $W\approx a f_1(x) + B \left| f_1(x)\right|$ has the same sign at $x_L \mbox{ and }x_R$; unbroken SUSY requires $a >B$ so that these signs are opposite. Thus, when we swap the ordered set of parameters $\{a,B\} \rightarrow \{ B+\hbar/2, a+\hbar/2\}$, the system goes through a change of supersymmetric phase, for $a\ne B$. For the case $a=B$, we note that neither $W$ nor $W'$ can be zero, since these would require $f_1=-\sqrt{f_1^2-\lambda}$, which is not possible for $\lambda\ne0.$ Therefore $(W^2)'=2WW'$ cannot be zero, so $W^2$ has no minimum and  there is only one intersection point. 
	
	\item \label{Enmurate:lambda>0B} $\lambda>0$ and $f_1^2 < \lambda$. 
	
	In this case $f_1$ is bound between $\pm \sqrt{\lambda}$. The derivative $f_1' = f_1^2 - \lambda <0$ approaches zero only as $f_1\to\pm\sqrt{\lambda}$  so $f_1$ asymptotically approaches $\pm\sqrt{\lambda}$ at $x_L$ and $x_R$ which must be $\pm\infty$, respectively \footnote{Since $\int_{x_0}^{\pm \infty} W(x) dx$ must be infinity for the normalizability of the groundstate, and because $W$ approaches a constant value at both ends of the domain, the domain of $x$ must be infinite.}.	
	Similar to the case of item (\ref{Enmurate:lamda<0}) above, the derivative $f_1$ is negative at point $x_0$ where  $f_1=0$.  All even derivatives are zero and odd derivatives are non-zero at this point. Thus, the resulting function $f_1$ is antisymmetric about $x_0$. Since $f_1$ is an odd function and $f_2$ vanishes at both ends of domain, the superpotential does not go into a broken supersymmetric phase for any value of the parameter.

	\item \label{Enmurate:lambda>0A} $\lambda>0$ and $f_1^2 > \lambda$. 

	In this case $f_1>\sqrt{\lambda}$ or $f_1<-\sqrt{\lambda}$.  
	If $f_1 > 0$ and $a B >0$, then $W^2$ has no minimum and there is only one intersection point. If $f_1>0$, $a B < 0$, and $ a\ne -B$, then SUSY cannot be broken. If $f_1>0$ and $a = -B$, then similar to the case $a=B$ for $\lambda <0$ above, there is only one intersection point. 
	
	If $f_1<0$ and \{$a B <0 $ or $a = B$\} then $W^2$ has no minimum and the BSWKB condition does not apply. If $aB > 0$, then SUSY is unbroken for $a> B$  and broken for $a < B$. Therefore the transformation $a \rightarrow B+\hbar/2, B\rightarrow a+\hbar/2$ changes from broken to unbroken SUSY. 
		
\end{enumerate}

\end{document}